\RequirePackage{fix-cm}
\PassOptionsToPackage{sort&compress}{natbib}
\documentclass{article}
\usepackage[preprint]{neurips_2018}

\usepackage[english]{babel}
\usepackage{blindtext}
\usepackage{multirow}

\usepackage{stackengine}

\usepackage{algorithm,algpseudocode}
\usepackage{tabto}
\usepackage{url}

\usepackage{graphicx}
\usepackage{lscape}
\usepackage[caption=false,labelformat=simple]{subfig}

\usepackage{latexsym}
\usepackage[export]{adjustbox}
\usepackage[strings]{underscore}
\setcitestyle{square,comma,numbers}

\usepackage{amsmath}
\usepackage{nccmath}
\usepackage{bm}

\usepackage{lscape}
\usepackage{longtable}
\usepackage{multirow}
\usepackage{booktabs}

\usepackage{color,soul}

\usepackage{array}
\newcolumntype{P}[1]{>{\centering\arraybackslash}p{#1}}

\begin{document}
\title{Engineering problems in machine learning systems}

\author{%
  Hiroshi Kuwajima \\
  DENSO CORPORATION \& Tokyo Institute of Technology\\
  \texttt{hiroshi.kuwajima.j7d@jp.denso.com, kuwajima@ok.sc.e.titech.ac.jp} \\
  \And
  Hirotoshi Yasuoka \\
  DENSO CORPORATION \\
  \texttt{hirotoshi.yasuoka.j2z@jp.denso.com} \\
  \And
  Toshihiro Nakae \\
  DENSO CORPORATION \\
  \texttt{toshihiro.nakae.j8z@jp.denso.com} \\
}

\maketitle


\begin{abstract}
Fatal accidents are a major issue hindering the wide acceptance of safety-critical systems that employ machine learning 
and deep learning models, such as automated driving vehicles. 
In order to use machine learning in a safety-critical system, 
it is necessary to demonstrate the safety and security of the system through engineering processes.
However, thus far, no such widely accepted engineering concepts or frameworks have been established for these systems. 
The key to using a machine learning model in a deductively engineered system
is decomposing the data-driven training of machine learning models into requirement, design, and verification, 
particularly for machine learning models used in safety-critical systems. 
Simultaneously, open problems and relevant technical fields are not organized in a manner 
that enables researchers to select a theme and work on it. 
In this study, we identify, classify, and explore the open problems in engineering (safety-critical) 
machine learning systems --- that is, in terms of requirement, design, and verification of machine learning models and systems --- as well as 
discuss related works and research directions, using automated driving vehicles as an example.
Our results show that machine learning models are characterized by a lack of requirements specification, 
lack of design specification, lack of interpretability, and lack of robustness. 
We also perform a gap analysis on a conventional system quality standard SQuARE with the characteristics of machine learning models 
to study quality models for machine learning systems.
We find that a lack of requirements specification and lack of robustness 
have the greatest impact on conventional quality models.
\end{abstract}

\section{Introduction}
Recent developments in machine learning techniques, such as deep neural networks (NNs), 
have led to the widespread application of systems that assign advanced environmental perception and 
decision-making to computer logics learned from big data instead of manually built rule-based logics~\cite{mlsys}. 
Highly complex machine learning techniques such as NNs have been studied for decades, however until recently, 
we have suffered from numerous training data to train complex models properly and computing methods 
to perform high computational complexity of training such models. 
The availability of big data and affordable high-performance computing, such as deep learning frameworks 
on off-the-shelf graphics processing units (GPUs)~\cite{Jia:2014:CCA:2647868.2654889}, 
have made highly complex machine learning techniques practical for various applications 
including automatic speech recognition~\cite{DBLP:conf/asru/GravesJM13}, image recognition~\cite{NIPS2012-4824}, 
natural language processing~\cite{NIPS2014-5346}, drag discovery~\cite{Vamathevan2019}, 
and recommendation systems~\cite{Elkahky:2015:MDL:2736277.2741667}. 

Machine learning models are becoming indispensable components even of systems 
that require safety-critical environmental perception and decision-making such as automated-driving systems~\cite{ml4its}. 
A safety-critical systems is a system whose failure may result in safety issues such as death or serious injury to people. 
For safety-critical systems, worst-case performance is more important than average performance, 
and developers are held strictly accountable.
However, for human society to accept such safety-critical machine learning systems, 
it is important to develop common engineering frameworks, such as quality measures and standard engineering processes, 
to manage the risks of using machine learning models and systems that include machine learning models~\cite{2016-01-0128}. 
Such frameworks, and ultimately the quality assurance based on them, have an impact on social receptivity 
because they can be one of the approaches used to deliver safety and security. 
In fact, recent accidents caused during the use of several experimental automated vehicles have revealed 
the imperative need to address the upcoming social issue of (quality) assurance based on such frameworks~\cite{dmv}. 
Engineering frameworks such as standard development processes have been studied for conventional systems and software for years, 
and machine learning systems also need such frameworks that engineers can follow. 
In order to establish engineering frameworks, it is necessary to visualize and organize these open problems; 
thus, experts from numerous different technical fields discuss these problems in depth and develop solutions driven by engineering needs. 

In this study, we review the open engineering problems associated with safety-critical machine learning systems 
and also present related works and future directions for research. 
We hypothesize an ideal training process that connects deductive requirements and data-driven training 
by considering test data as a requirements specification and training data as a design specification; 
thereafter, we review open problems for the process. 
Our results show that machine learning models are characterized by a lack of requirements specification, 
lack of design specification, lack of interpretability, and lack of robustness. 
In addition, we discover that requirements specification and verification for open environments are key aspects of machine learning systems.
We also study quality models for machine learning systems, 
which can be used for future requirements and evaluations of these machine learning systems.
Our results show that a lack of requirements specification and lack of robustness 
have the greatest impact on conventional system quality models.

\section{Background}
An automated driving vehicle is a vehicle that operates without human input. 
Automated driving has not been built as a stand-alone system in a vehicle but can be realized 
using a system comprising clouds, roadside devices (fog or cloud edge), 
and automated driving vehicles (edge)~\cite{app8020303}, 
which create and update high-precision digital maps~\cite{DBLP:conf/itsc/PoggenhansPJONK18} 
while cooperating with peripheral vehicles. 
An in-vehicle automated driving system installed in a vehicle comprises multiple subsystems 
for perception, planning, and control; 
such a system realizes automated driving operations in cooperation with 
clouds and roadside units~\cite{DBLP:conf/itnac/AliC11}. 
For simplicity, in this paper, we focus on these in-vehicle automated driving systems.
Each perception, planning, and control subsystem may contain necessary machine learning models. 
Supervised learning models~\cite{NIPS2012-4824} 
and reinforcement learning models~\cite{DBLP:journals/corr/MnihKSGAWR13} can be used 
for perception and planning, while non-machine learning control algorithms can be used for control. 
In order to build a machine learning system, it is necessary to define its engineering processes 
and quality measures in advance, then follow and measure them strictly during development time.
Conventional systems were developed in a rigorous development process involving requirement, design, 
and verification, cf. V-Model~\cite{INCOSE} (a graphical representation of a systems development lifecycle).

\begin{figure}[htbp]
  \centering
  \includegraphics[scale=0.4]{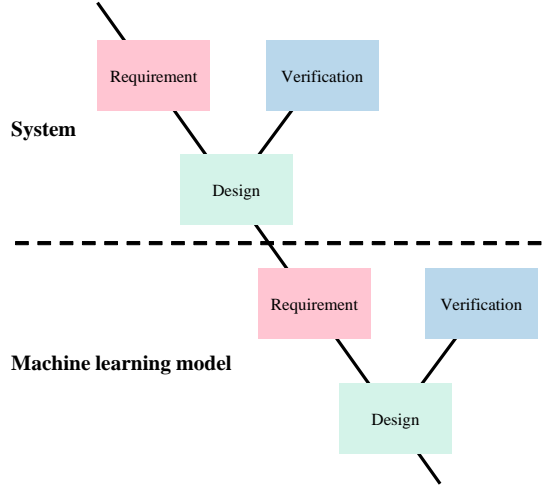}
  \caption{Engineering process of machine learning systems}
  \label{fig:mlse_process}
\end{figure}

In this study, we identify open engineering problems at two levels --- systems and machine learning models --- and 
use an automated driving system as an example of a safety-critical machine learning system. 
We proceed to investigate the problems in terms of the three steps of the development process: 
requirement, design, and verification.
The two levels and three steps are illustrated in Fig.~\ref{fig:mlse_process}. 
Notably, many of the problems considered in this paper do not occur only in automated driving systems 
but also generally in safety-critical systems.

This study is related to preceding studies~\cite{DBLP:journals/corr/abs-1709-02435,10.1007/978-3-319-67383-7-21} 
that studied the applicability of ISO 26262~\cite{ISO26262} and Automotive SPICE~\cite{ASPICE} to 
automotive software using machine learning and deep learning. 
Our work assumes a more general development process to show open problems; we examined quality models 
for machine learning systems, based on a conventional system and software quality standard, 
Systems and software Quality Requirements and Evaluation (SQuARE)~\cite{iso}, 
which has not been done in previous studies.

\section{Engineering machine learning models}
A machine learning model is acquired by executing a training algorithm with a model structure and training data sets for inputs, 
while trained models are evaluated using test datasets~\cite{murphy2013machine}. 
This is a data-driven inductive method that differs from the deductive development used for conventional systems. 
In this paper, we call a machine learning model that has undefined parameters a "model structure." 
In order to use machine learning models in a deductively engineered system, 
it is necessary to break down the data-driven training of model parameters into requirements, designs, and verifications, 
particularly for models used in safety-critical systems.

\begin{figure*}[htbp]
  \centering
    \subfloat[Conventional training process]{
    \centering
    \includegraphics[scale=0.4]{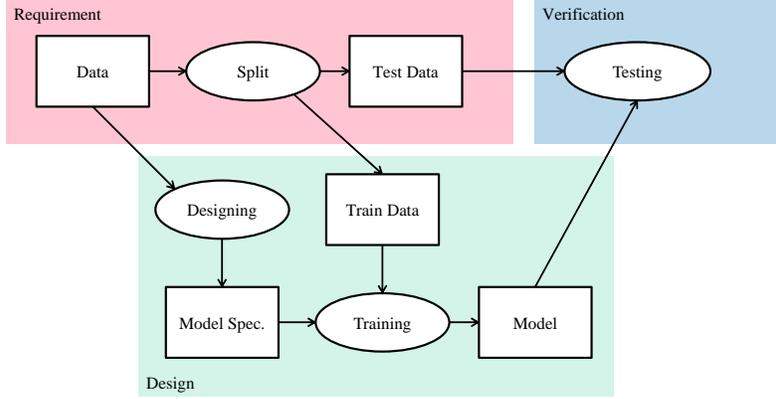} 
      } \\
    \subfloat[Hypothesized training process]{
    \centering
    \includegraphics[scale=0.4]{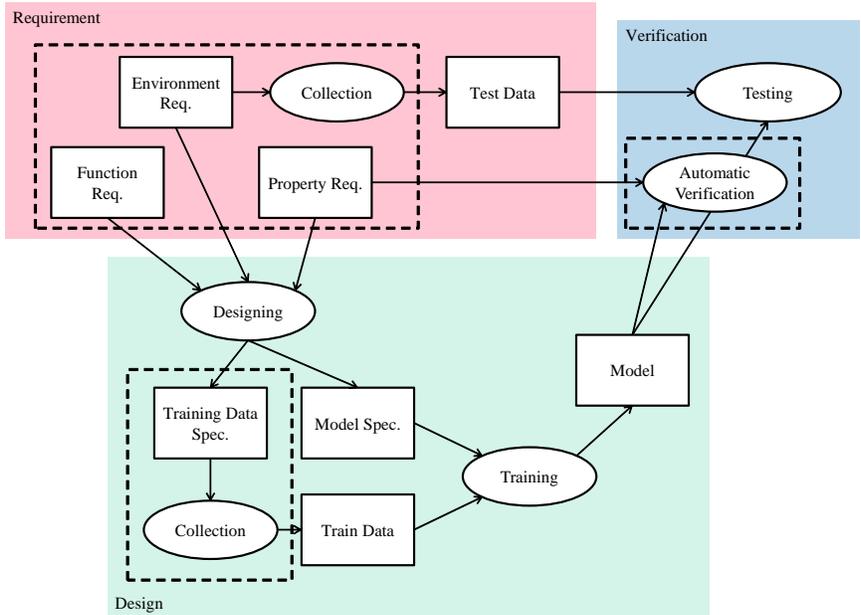} 
      }
  \caption{Engineering process of machine learning models}
  \label{fig:mlse_model}
\end{figure*}

We hypothesize the engineering process for machine learning models in Fig.~\ref{fig:mlse_model}. 
The dotted boxes in the figure illustrate the differences between the conventional training process 
and hypothesized training process.
A requirement of machine learning models can be the specification of test data, 
although the current practice is to divide the original data into training and test data sets~\cite{stone1974cross,arlot2010survey}. 
The design process then specifies or builds the training data to achieve high performance in the test data, 
with the model requirements as a background.
The explicit specification of test data and training data addresses a lack of requirements specification 
and a lack of design specification, respectively.
In the current practice, the verification of machine learning models is measured using performance metrics on the test data. 
However, we consider it important to check properties that cannot be measured using the test data, 
such as robustness and interpretability. 

In the following subsections, we introduce our ideas related to 
the requirements, designs, and verifications of machine learning models, as well as research directions and related works.

\subsection{Requirements of machine learning models}
Most current machine learning research 
undoubtedly assumes that test data is given~\cite{NIPS2012-4824,He2016DeepRL}; 
it is the main part of a model'{}s requirements. 
Test data must be carefully specified at the beginning of development, 
by either the developers or contractees of the machine learning model, and must be agreed upon by their contractors. 
Thus, the main open engineering problem here is the deductive definition of the requirements for machine learning models 
and their test data to enable the test data to connect with deductive requirements and data-driven training.
In machine learning, the roles of training data and test data must be considered to be different. 
While training data is used to improve the performance of a machine learning model~\cite{James:2014:ISL:2517747}, 
we propose considering the test data to accurately reflect the environmental conditions in operation. 
However, in practice~\cite{DBLP:journals/ml/Ben-DavidKM97}, when all data obtained at the time of development are divided, 
some are used as training data and the others become test data~\cite{stone1974cross,arlot2010survey}. 
For simplicity, we ignore validation data for model selection. 
Despite the ultimate goal of machine learning models to work well in operation, we test machine learning models on test data, 
which originates from the same source as training data.
In this manner, the training and test data are approximately equally distributed, 
but their relationship to the operational data (which is the actual target of the model) is unknown 
or it is implicitly assumed that the training, test, and operational data sets are similar ~\cite{DBLP:journals/jmlr/BickelBS09}.
In other words, machine learning models are trained using data-driven methods that {\bf lack requirements specification}.

In particular, in a safety-critical machine learning system, it is necessary to specify the distribution of test data 
(considering the operational environment the system will actually be operated in) 
and to collect the test data based on these specifications. 
By accepting an a priori viewpoint of the distribution of test data, 
we can define the assumed environment deductively and collect data inductively. 
Moreover, by assuming the distribution of test data, we can discuss the operational domain 
(operational data distribution) for requirements specification.
Operational data tend to change with time, 
thereby deteriorating model performance in operation~\cite{Tsymbal04theproblem,Webb:2016:CCD:2962863.2962874}. 
The deviation between test data used during development and operational data can become larger with time 
from what it was when development was completed.
This phenomenon is referred to either as covariance shift~\cite{DBLP:journals/jmlr/BickelBS09}, 
distributional shift~\cite{DBLP:journals/corr/AmodeiOSCSM16}, or concept drift~\cite{Tsymbal04theproblem}.
If the operational data trend changes from that of test data, then machine learning models trained on the test data 
do not work on the changed operational data.
Thus, it is important to check for consistency between operational data and test data (assuming the original environment) 
and to either make the machine learning models follow the operational data in a continuous maintenance process or to, at least, 
detect the deviation between test and operational data. A lack of requirements specification is a barrier to this.

\subsubsection*{Related works and research directions for requirements of machine learning models}
Although it does not incorporate the specification of test data, {\it i.e.}, requirements specification, 
runtime monitoring of neuron activation patterns is an approach to detect change points ~\cite{DBLP:journals/corr/abs-1809-06573}. 
It creates a monitor of neuron activation patterns after training time, 
and runs the monitor at operation time to measure the deviation from training time. 
Change is detected when the activation pattern at operation time becomes detached from the neuron activation pattern at training time.
Neuron activation patterns on test data may implicitly include the model requirements as a background.

Even in the current development of in-vehicle automated driving systems, 
the test data would be collected assuming the operational environment, 
in order to make the distribution of the operational data and that of the test data as consistent as possible. 
However, the methods used to describe the assumed environment of machine learning models are not organized. In particular, 
specific methods are required to define the completeness of test data. 
In previous literature, CV-HAZOP (Computer Vision - Hazard and Operability Studies)~\cite{DBLP:conf/iccv/ZendelMHH15} 
defined a catalogue of challenges (risks) for computer vision algorithms. 
The catalogue has $1,469$ manually registered risks as of now. 
Including all CV-HAZOP risks can be a test data coverage in computer vision problems. 
When systematically testing machine learning models to achieve test data coverage, 
we experience combinatorial explosion while guiding the data sampling process. 
In previous literature, 
quantitative projection coverage was used to resolve such combinatorial explosion~\cite{DBLP:conf/atva/ChengHY18}.

Although these previous works focused on combinatory environments, the importance or criticality of each environment could change. 
For example, criticality of misclassification of pedestrians may be high in daytime city street, 
whereas that of vehicles may be high in night time highway.
CV-HAZOP proposes that the catalogue of challenges creates a basis for 
referencing criticalities for each risk and calculating criticality coverage~\cite{DBLP:conf/iccv/ZendelMHH15}. 
Figure~\ref{req_spec} illustrates our proposed example of requirements specification.
Test data must have attributes, such as time and weather, 
and their distributions that are based on the assumed environment (Fig.~\ref{env_spec}). 
Recent public driving data sets have such attributes. 
For example, BDD100K~\cite{DBLP:journals/corr/abs-1805-04687} has 
weather conditions, including sunny, overcast, and rainy, different times of day, 
including daytime and nighttime, as well as scenes, including city street, gas stations, highway, parking lot, residential, and tunnel. 
Further, since the required performance may change for each environment, 
it is necessary to express the association between the assumed environment and the required performance. 
Each condition of the test data distribution can have a different confusion matrix (or other performance metrics) 
that machine learning models will have as desired values (Fig.~\ref{perf_spec}). 
\begin{figure*}[h]
  \centering
    \subfloat[Environment specification]{
  \label{env_spec}
    \centering
  \begin{tabular}{cc|c|c|}\cline{3-4}
     &  & \multicolumn{2}{c|}{time} \\ \cline{3-4}
     &  & day & night \\ \hline
  \multicolumn{1}{|c|}{\multirow{2}{*}{weather}} & fine  & 40\% & 30\% \\ \cline{2-4}
  \multicolumn{1}{|c|}{\multirow{2}{*}{}} & rainy  & 20\% & 10\% \\ \hline
  \end{tabular} 
  } \\
    \subfloat[Performance specification for fine $\times$ day $\dagger$]{
    \centering
  \begin{tabular}{cc|c|c|}\cline{3-4}
& & \multicolumn{2}{c|}{predicted} \\ \cline{3-4}
     &   & pedestrian & vehicle \\ \hline
  \multicolumn{1}{|c|}{\multirow{2}{*}{actual}} & pedestrian  & 90\% & 10\% \\ \cline{2-4}
  \multicolumn{1}{|c|}{} & vehicle  & 20\% & 80\% \\ \hline
  \end{tabular}
  }
    \subfloat[Performance specification for fine $\times$ night]{
  \label{perf_spec}
    \centering
  \begin{tabular}{cc|c|c|}\cline{3-4}
& & \multicolumn{2}{c|}{predicted} \\ \cline{3-4}
     &   & pedestrian & vehicle \\ \hline
  \multicolumn{1}{|c|}{\multirow{2}{*}{actual}} & pedestrian  & 85\% & 15\% \\ \cline{2-4}
  \multicolumn{1}{|c|}{} & vehicle  & 15\% & 85\% \\ \hline
  \end{tabular}
  } \\
    \subfloat[Performance specification for rainy $\times$ day]{
    \centering
  \begin{tabular}{cc|c|c|}\cline{3-4}
& & \multicolumn{2}{c|}{predicted} \\ \cline{3-4}
     &   & pedestrian & vehicle \\ \hline
  \multicolumn{1}{|c|}{\multirow{2}{*}{actual}} & pedestrian  & 90\% & 10\% \\ \cline{2-4}
  \multicolumn{1}{|c|}{} & vehicle  & 20\% & 80\% \\ \hline
  \end{tabular}
  }
    \subfloat[Performance specification for rainy $\times$ night $\ddagger$]{
    \centering
  \begin{tabular}{cc|c|c|}\cline{3-4}
& & \multicolumn{2}{c|}{predicted} \\ \cline{3-4}
     &   & pedestrian & vehicle \\ \hline
  \multicolumn{1}{|c|}{\multirow{2}{*}{actual}} & pedestrian  & 85\% & 15\% \\ \cline{2-4}
  \multicolumn{1}{|c|}{} & vehicle  & 15\% & 85\% \\ \hline
  \end{tabular} 
  } \\ \vspace{2mm}
  \raggedright
  $\dagger$ E.g., there are many pedestrians in fine daytime, and they are prioritized. \\
  $\ddagger$ E.g., there are many vehicles in rainy nights, and they are prioritized.
  \caption{Example environment requirements specification (data distribution matrix) and performance requirements specification (confusion matrix)}
  \label{req_spec}
\end{figure*}

\subsection{Design of machine learning models}
A machine learning model is automatically obtained by training the parameters of a model structure using training data. 
Thus, specifications cannot be designed a priori --- that is, machine learning models {\bf lack design specifications}). 
This limitation is essential and unavoidable because high-performance machine learning models 
are developed by learning high-dimensional parameters from data that engineers cannot manually specify. 
However, in the development of a safety-critical machine learning system, 
it is necessary to record the model structure, training data, and training system, 
including training specifications --- such as hyper parameters, initial parameters, learning rates, 
and random number seeds --- to secure the reproducibility of the training process. 

Engineers cannot design the training; however, they can design the training data. 
Training data, as a large indirect part of the design specification, 
coupled with training specifications is carefully designed to achieve the requirements specification. 
In this manner, the lack of design specification is indirectly remedied. 
Yet, to the best of our knowledge, there is no standard or widely accepted process of designing training data 
for machine learning models.

\subsubsection*{Related works and research directions for design of machine learning models}
One of the challenges with the lack of design specification is the establishment of a training process 
for machine learning models by designing training data and models. 
Training data must be designed in the process by iteratively identifying the weak points of the model 
and then generating or collecting additional data for training. 
A previous suggestion~\cite{andrewng} indicates that a criteria for growing training data is that the training error is low 
while the test error is high; 
however, the suggestion does not show what types of data must be added.
It is known that deep learning models, in particular, easily fit a random labeling of the training data~\cite{DBLP:journals/corr/ZhangBHRV16} 
and, thus, the distribution of training data is important.

\subsection{Verification of machine learning models}
Machine learning models are mainly verified by running a model on test data; 
however, certain properties of a machine learning model, such as robustness, cannot be evaluated with test data. 
Therefore, we introduce property checking in the verification of machine learning models.

An increasing stability against disturbance, or a {\bf lack of robustness}, is key to the verification of machine learning models. 
It has been reported that image recognition models incorrectly recognize slight noise 
that cannot be recognized by humans with high confidence, 
thereby creating what are called adversarial examples (AEs)~\cite{DBLP:journals/corr/SzegedyZSBEGF13}. 
An AE is known to have model-independent versatility and is an issue that can threaten the safety of automated driving systems, 
depending on image recognition. 
For example, when evaluating robustness against an AE as fault tolerance, 
it is necessary to artificially generate perturbations around data points.
We can generate an AE close to a data point specified in the requirements and quantify the robustness 
using the maximum radius in which the model can yield correct answers. 

The inference processes of advanced machine learning models --- such as NNs --- are considered 
black boxes, and machine learning models {\bf lack interpretability}. 
In this context, a black box refers to a situation where, although feature activations can be observed physically, 
the actual phenomenon cannot be understood. 
That being said, safety-critical systems must exhibit interpretability and transparency. 
The interpretability of machine learning models has been well-researched recently and there are several methods for addressing it.
LIME~\cite{lime} is one of the most well-known methods for improving interpretability. 
It derives a simple interpretable model to explain the behavior of an original model around a given data point. 
NN visualization~\cite{grun16featurevis} also shows great promise to improve interpretability. 
Object detectors emerging in deep scene CNNs is an NN visualization 
that intentionally performs occlusion on input data and specifies the region 
where the inference result changes drastically as a region of interest~\cite{scenecnn-iclr15,cvpr2016-zhou};
another method back-propagates activation values from the influencer nodes during the subsequent feature extraction process 
to identify the region of interest~\cite{10.1007/978-3-319-10590-1-53} and generate 
heat maps~\cite{DBLP:journals/corr/ShrikumarGSK16,DBLP:conf/icann/BinderMLMS16,DBLP:journals/pr/MontavonLBSM17,simonyan14deep} 
for convolutional NNs.
Further, interpretability is also useful for performance improvement, debugging during training, and validating of training results. 
Developers can understand the internal behavior of a trained NN to train higher performance models~\cite{Kuwajima2019}. 
For example, a developer can visualize an NN's focus points for an incorrect inference 
and understand what was wrong, before additional training data is collected according to the analysis. 
If a machine learning model outputs an incorrect inference, but the visualized focus area is natural for humans, 
then an inaccurate ground truth label is suggested.

\subsubsection*{Related works and research directions for verification of machine learning models}
In the field of theoretical computer science, 
the automatic design verification~\cite{Lam:2008:HDV:1522514} based on formal verification technologies 
for certain properties, such as safety and liveness~\cite{Garcia2006FormalVO}, 
makes the verification of a machine learning model possible. 
Several automatic verification techniques exist for NNs and we categorize them here. 
The initial categories are function and decision problems. 
The former quantifies the degrees of properties, while the latter identifies if the properties are satisfied in a machine learning model. 
Related works for function problems address adversarial frequency and severity~\cite{NIPS2016-6339} 
as well as maximum perturbation bound~\cite{Cheng2017MaximumRO}, referring to the frequency of AE found, 
the expectation of the closest AE, 
and the maximum absolute value of the perturbation of inputs that do not change the outputs, respectively. 
Decision problems are further subdivided into verification and falsification, 
which seek a complete proof and counterexamples by best effort, respectively. 
Related works of verification are global safety~\cite{Pulina:2012:CSS:2350156.2350160}, 
local safety~\cite{Pulina2010AnAA}, $(\epsilon, \delta)$-robustness~\cite{KBD+17-FVAV}, 
and $(x, \eta, \delta)$-safe~\cite{Huang2017SafetyVO}.
Global safety is output bound, and local safety is the consistency of inference among close data points.
A related example for falsification is the CNN Analyzer~\cite{Dreossi,dreossi-rmlw17}. 
It identifies counterexamples against the signal temporal logic~\cite{donze2013signal} properties of in-vehicle automated driving systems 
and counterexamples of object (vehicle) detection by convolutional NNs.
Further, Reluplex~\cite{Katz2017ReluplexAE} is a solver used to both verify and 
falsify first-order propositional logics~\cite{Andrews:2002:IML:581793} 
against NNs using Rectified Linear Units (ReLU)~\cite{DBLP:conf/icml/NairH10} for activation functions. 
Reluplex is an SMT (satisfiability modulo theories) solver~\cite{DeMoura:2008:ZES:1792734.1792766} to verify properties of deep NNs 
or provide counterexamples against them by utilizing the simplex method~\cite{dantzig1987origins} 
and the partial linearity of the ReLU function. 
Dependability metrics set for NNs is a related work that proposes metrics such as 
scenario coverage, neuron activation pattern, interpretation precision 
for RICC (robustness, interpretability, completeness, and correctness) criteria~\cite{DBLP:conf/memocode/ChengNHRY18}. 

\section{Engineering machine learning systems}
In this section, we review open engineering problems in terms of the system level of in-vehicle automated driving systems 
as an example of safety-critical machine learning systems. 
Problems related to machine learning systems originate from machine learning models and the open environments 
in which automated vehicles function. 
The former is low modularity of machine learning systems due to the characteristics of machine learning models, 
such as lack of design specifications and lack of robustness. 
The latter include capturing physical operational environments and user behaviors of in-vehicle automated driving systems for requirements 
and addressing the intractableness of field operation testing (FOT) for verification. 
An open environment problem is not directly related to machine learning, although it is an important challenge 
for in-vehicle automated driving systems. 
In this paper, we consider open environments to be a common challenge for machine learning systems 
because machine learning models are employed to capture these complex environments.

\subsection{Requirements of machine learning systems}
In order to develop high quality systems and products, comprehensive requirements specifications and 
the evaluation of machine learning systems based on the requirements specification are needed; 
in turn, these require appropriate quality characteristics for the systems that can be used for requirements and evaluations. 
Quality characteristics of machine learning "systems" are more important in the context of industry than those of machine learning "models," 
because machine learning models are not used in a stand-alone manner but are always embedded in systems.
System and software quality models have been developed for years; 
however, to the best of our knowledge, there is no standard quality model that adapts the characteristics of 
machine learning models --- such as lack of requirements specifications, design specifications, interpretability, and robustness --- into account.
Thus, we conduct a gap analysis on a conventional system and software quality standard, 
SQuARE~\cite{iso} in \ref{quality}.

Another important aspect of machine learning (or any other) systems is that they cannot operate in every environment 
and require limitations or warranty scopes. 
Thus, a particular machine learning system must be implemented for a predefined environment. 
Environment attributes to be predefined for automated driving systems are static conditions such as weather, times of day, scene, 
road, as well as dynamic conditions (dynamics of) such as the vehicle under control and other moving objects 
(surrounding vehicles and pedestrians).
However, there are various (uncountable) types of roads, traffic lights, and traffic participants, 
such as other vehicles (be they automated or manually driven) and pedestrians; 
therefore, it is not easy to define the operational environment for in-vehicle automated driving systems.
An open engineering problem in the requirements specification of machine learning systems is 
that there is no standard means to design and define such environments, {\it i.e.}, requirements specification cannot be clearly defined.
In the automotive industry, this is called the operational design domain~\cite{ADS2,AV3} and it can be defined by conditions 
such as geographical areas, road types, traffic conditions, and maximum speed of the subject vehicle~\cite{ODD}.

\subsubsection*{Related works and research directions for requirements of machine learning systems}
The German PEGASUS project is a joint initiative of vehicle manufacturers, suppliers, tool vendors, certification organizations, 
and research institutes, 
aiming to define standard quality assurance methods for automated-driving systems~\cite{pegasus}. 
The purpose of this project is to clarify the expected performance level and evaluation criteria of automated driving systems 
through scenario-based verification. 
The scope of the project includes standard test procedures, continuous and flexible tool chains, 
the integration of tests into development processes, 
cross-company test methods, requirement definition methods, driving scenarios, and a common database of scenarios.
Scenarios are collected from test drives and the market to demonstrate that systems are equal to, or better than, human drivers. 
Scenario collection, {\it i.e.}, building requirements specification, and scenario-based verification are conducted in a continuous manner. 
Regular scenarios are continuously tested by simulation, 
and critical scenarios are tested through artificially configured environments on test courses. 
The PEGASUS project is an excellent example of the continuous requirements and verification 
for in-vehicle automated driving systems and their verification.

\subsection{Design of machine learning systems}
An open engineering problem at the system level of machine learning systems is designing systems that include 
machine learning models by considering and applying the characteristics of 
"Change Anything Change Everything" (CACE)~\cite{Sculley:2015:HTD:2969442.2969519}.
CACE originates from a lack of design specification in machine learning models. 
Machine learning models are trained in a data-driven manner, thereby making the localizing of change difficult. 
If a small part is changed, then the entire machine learning changes once it is trained again. 
Subsequently, machine learning systems have to be changed for the newly trained machine learning models. 
In order to prevent reworking after training machine learning models, 
it is necessary to have system architectures that can cope with additional requirements without modification of the model.

In general, it is difficult for a machine learning model to achieve 100\% accuracy on test data~\cite{Domingos:2012:FUT:2347736.2347755} 
and as its accuracy approaches 100\%, further performance improvement becomes difficult. 
Therefore, optimizing machine learning models is not the only means to improve subsystem performance, 
thereby making a rigorous breakdown of subsystem requirements into machine learning model requirements essential 
for safety-critical machine learning systems. 
In this process, safety analysis methods and processes are important, 
such as the encapsulation of machine learning models by rule-based safeguarding and the use of redundant and diverse architecture 
that absorbs and mitigates the uncertainty of machine learning models.

\subsubsection*{Related works and research directions for design of machine learning systems}
To the best of our knowledge, we do not find special techniques that directly address the design of machine learning systems.
SOTIF~\cite{sotif-anal}, a safety standard/process concerning performance limits of functionalities, 
focuses on securing functionalities with uncertainty.
Uncertain functionalities include machine learning models.
SOTIF has a process that includes identification of scenarios that can trigger unsafe actions (triggering conditions) 
for the system and system modifications to address them~\cite{SafetyAnalysisMethods}.
The process standards can be potentially effective in the design of machine learning system in general, 
rather than evaluating an entire machine learning system upon completion of development.
In addition to process standards, research directions include test stubs for machine learning models, 
encapsulation of machine learning models by rule-based safeguarding, 
and the use of redundant and diverse architecture that mitigates and absorbs the low robustness of machine learning models. 

\subsection{Verification of machine learning systems}
The simplest approach to verifying an in-vehicle automated driving system is by verification against actual data. 
Accumulating a large number of safe automated driving trips, with long distances to match human drivers, 
will effectively demonstrate that in-vehicle automated driving systems are as safe as human drivers. 
In order to verify the system within a realistic time-frame, there are two options: 
reduce the required verification scenarios or accelerate the verification. 
Therefore, high accuracy verification models must be able to exclude unreal scenarios. 
It is necessary to accelerate simulation experimentation, thereby reproducing corner-case scenarios on test courses 
with a short mileage ({\it i.e.}, 
scenarios with an extremely low probability of occurrence and ones that are difficult to statistically obtain through FOT on an actual road).

\subsubsection*{Related works and research directions for verification of machine learning systems}
Obtaining statistically significant results would require FOT on a humongous number of miles~\cite{KALRA2016182}. 
\cite{KALRA2016182} is based on a simple hypothesis testing, and the resulting required miles may not reflect actual situations. 
Research directions include building detailed close-to-reality models for driving scenes and scenarios 
to reflect the real world conditions and reduce FOT miles.

\section{Quality of machine learning systems}
\label{quality}
We reviewed the open engineering problems in machine learning systems, 
and recognized that machine learning models are characterized by 
their lack of requirements specifications, design specifications, interpretability, and robustness. 
In this section, we study quality models for machine learning systems 
by discussing the combination of these machine learning characteristics 
and a conventional system and software quality standard, SQuARE\cite{iso}. 

\subsection{Quality models for conventional systems}
\begin{figure*}[htbp]
  \centering
  \includegraphics[scale=0.37]{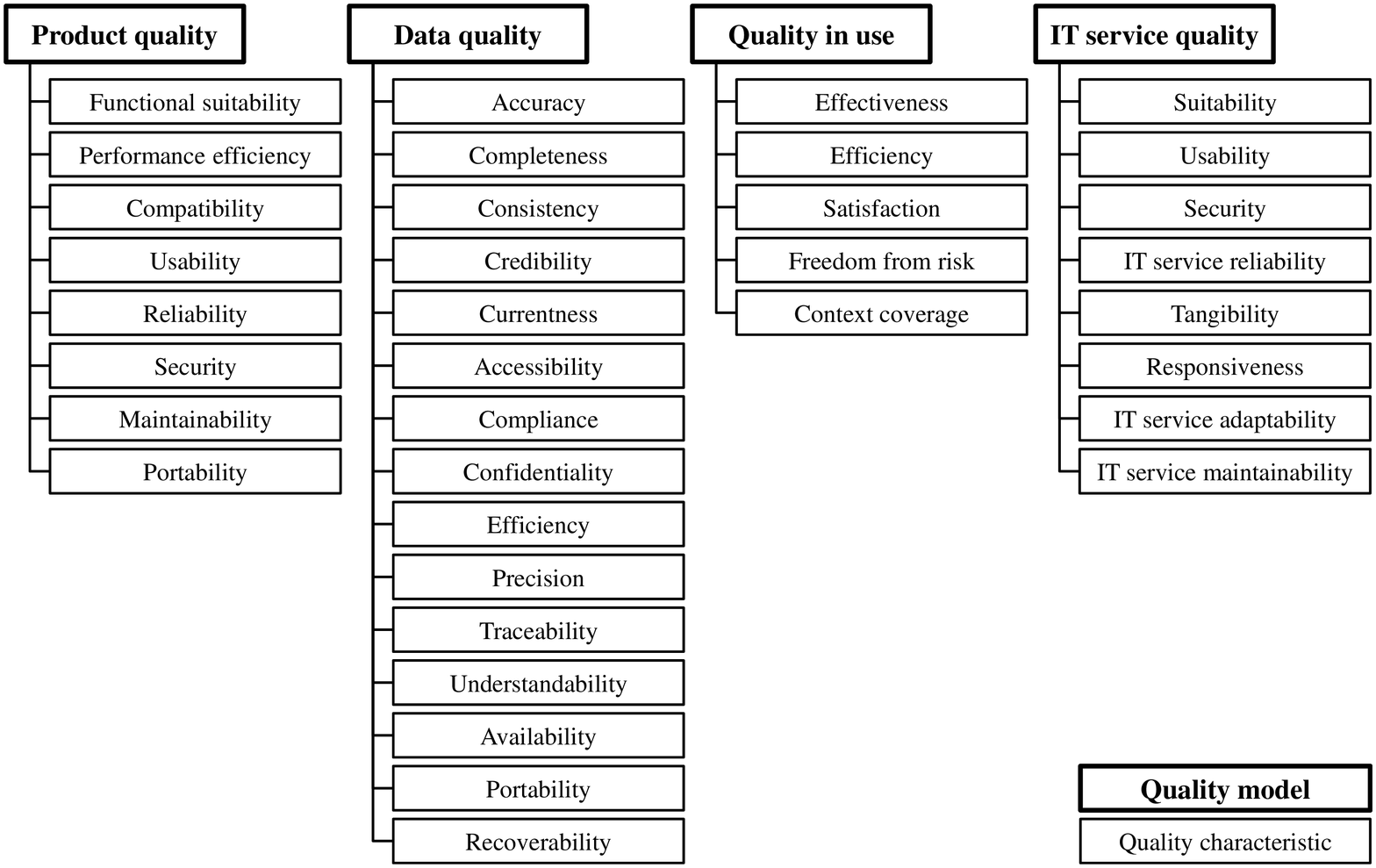}
  \caption{Quality models and quality characteristics in SQuARE}
  \label{fig:q_model}
\end{figure*}

We focus on SQuARE, ISO/IEC 25000 series~\cite{iso}, as the conventional system quality baseline. 
Systems and software quality are usually studied in software engineering.
One of the earliest work is an international standard 
ISO/IEC 9126 Software engineering --- Product quality~\cite{ISO9126}, first issued in 1991.
ISO/IEC 9126 classified software quality into six characteristics: Functionality, Reliability, Usability, Efficiency, Maintainability, and Portability. 
ISO/IEC 9126 was replaced by its succeeding international standard ISO/IEC 25000 series, 
Systems and software engineering --- Systems and software Quality Requirements and Evaluation (SQuaRE)~\cite{iso}.
SQuARE is a widely acknowledged system quality standard and includes 
quality measures (QMs) and quality measure elements (QMEs) as well as quality models, characteristics, and sub-characteristics. 
These components have a tree structure (one-to-many relationships), and the top-level quality models are 
Product quality, Data quality, Quality in use, and IT service quality, as illustrated in Fig.~\ref{fig:q_model}. 
Boxes with thick lines and thin lines in Fig.~\ref{fig:q_model} represent quality models and quality characteristics, respectively. 
Quality sub-characteristics are not defined for Data quality.

Each quality characteristic of Data quality, or each quality sub-characteristic of Product quality and Quality in use, has multiple QMs 
that define how to quantify the quality. A QM $X$ is defined in the form of a formula, such that $X = A / B$ and $X = 1 - A / B$, 
and the elements in the formula $A$ and $B$ are QMEs. 
An example set of a quality model, a characteristic, a sub-characteristic, a QM, and QMEs are 
Product quality, Reliability, Maturity, Mean time between failure, 
and Operation time (QME 1) and Number of system/software failures that actually occurred (QME 2), respectively. 
There are other QMs for the sub-characteristic Maturity (such as Failure rate, whose QMEs are 
Number of failures detected during observation time and Duration of observation).
QMs and QMEs are not defined for IT service quality. 

\subsection{Gap analysis}
We performed a gap analysis between conventional system quality models and future system quality models for machine learning systems, 
given a conventional system and software quality standard SQuARE 
and the characteristics of the machine learning models introduced in this paper. 
In order to conduct the most fine and precise analysis, 
we checked each QME (such as the number of systems/software failures that actually occurred) 
against each machine learning characteristic (such as a lack of robustness) 
to see if the QME was affected by the machine learning characteristic. 
If a QME in machine learning systems became immeasurable, 
as is the case with conventional systems, then the parent quality (sub-)characteristic would have gaps. 
IT service quality model was ignored in this gap analysis because it has no QME defined in the ISO/IEC 25000 series.
Table~\ref{qme_sample} presents an example of impact analysis of characteristics of machine learning models and QMEs defined for 
Functional suitability in Product quality. Req, Des, Rob, and Tra are abbreviations for 
lack of requirements specification, design specification, robustness, and transparency, respectively.
Functional suitability in Product quality was selected only to serve as an example, 
and corresponding analysis was conducted for all QMEs of all quality models, except for IT service quality model.
\begin{table*}[h]
\centering
\caption{Example impact analysis (Functional suitability characteristic in Product quality model)} \label{qme_sample}
\begin{tabular}{@{} lm{6cm}P{0.45cm}P{0.45cm}P{0.45cm}P{0.45cm} @{}} \toprule
QM & QME & Req & Des & Rob & Tra\\ \midrule
\multirow{2}{3.6cm}{Functional coverage} & Number of functions missing &  &  &  & \\ \cmidrule{2-6}
 & Number of functions specified &  &  &  & \\ \midrule
\multirow{2}{3.6cm}{Functional correctness} & Number of functions that are incorrect &  & $\dagger$ &  & \\ \cmidrule{2-6}
 & Number of functions considered &  &  & $\ddagger $ & \\ \midrule
\multirow{2}{3.6cm}{Functional appropriateness of usage objective} & Number of functions missing or incorrect among those that are required for achieving a specific usage objective. &  & $\dagger \dagger $ &  & \\ \cmidrule{2-6}
 & Number of functions required for achieving a specific usage objective &  &  & $\ddagger \ddagger $ & \\ \midrule
\multirow{3}{3.7cm}{Functional appropriateness of system} & Appropriateness score for a usage objective &  &  &  & \\ \cmidrule{2-6}
 & Number of usage objectives &  &  &  & \\ \bottomrule 
\end{tabular}\\~\\
  \raggedright
$\dagger$ When the input changes slightly, the result can changes drastically. We cannot measure the correctness of the function precisely. Perturbed trials can quantify the uncertainty.\\
$\ddagger$ Functions considered cannot be defined strictly. For example, there are many pedestrian variations of pedestrian detection for an auto emergency braking (AEB) function, and it can be multiple functions. We cannot define functions without ambiguity.\\
$\dagger \dagger $ When the input changes slightly, the result can changes drastically. We cannot measure the correctness of the function precisely. Perturbed trials can quantify the uncertainty.\\
$\ddagger \ddagger $ Functions considered cannot be defined strictly. For example, there are many pedestrian variations for pedestrian detection, and it can be multiple functions. We cannot define functions without ambiguity.
\end{table*}

We examined $1,464$ combinations of $366$ QMEs and $4$ characteristics of machine learning models to obtain the results. 
The number of combinations we identified as being affected by machine learning models was $20$ from among $1,464$. 
Tables~\ref{prod_q}, \ref{data_q}, and \ref{use_q} are the summaries of impact analysis on 
Product quality, Data quality, and Quality in use models affected by the characteristics of machine learning models. 
QM and QME levels are omitted.
Each QME associated with a quality (sub-)characteristic was examined to determine 
if it was affected by any machine learning characteristics: 
a lack of requirements specification (Req), a lack of design specification (Des), a lack of robustness (Rob), or and lack of transparency (Tra). 
The section signs with numbers in parentheses next to machine learning characteristics 
are the indices to the itemization in the subsequent paragraphs.
The number of QMEs affected by the machine learning characteristics are presented in the abovementioned tables. 
If we consider that the ratios of QMEs affected by characteristics of machine learning models 
are an indication of the impacts to quality (sub-)characteristics, 
then at the quality-model level, it is evident that the impact to Product quality is the highest, 
while those of Data quality and Quality in use are low. 
\begin{table*}[h]
\caption{Impact analysis on Product quality model} \label{prod_q}
\begin{tabular}{@{} llrrl @{}} \toprule
& & \multicolumn{3}{c}{Number of QMEs} \\ \cmidrule{3-5}
Characteristic & Sub-characteristic & all & \multicolumn{2}{c}{affected} \\ \midrule
Functional suitability & Functional correctness & $2$ & $2$ & Req (\S 5), Rob (\S 9)\\
& Functional appropriateness & $4$ & $2$ & Req (\S 6), Rob (\S 10)\\
& Others & $2$ & $0$\\ 
{\bf Subtotals} &  & $8$ & $4$\\ \midrule
Performance Efficiency & Any & $29$ & $0$\\
{\bf Subtotals} &  & $29$ & $0$\\ \midrule
Compatibility& Any & $8$ & $0$\\
{\bf Subtotals} &  & $8$ & $0$\\ \midrule
Usability & Operability & $18$ & $1$ & Tra (\S 19)\\
& Others & $25$ & $0$\\
{\bf Subtotals} &  & $43$ & $1$\\ \midrule
Reliability & Maturity & $8$ & $2$ & Rob$\times 2$ (\S 12, \S 13)\\
& Fault tolerance & $7$ & $2$ & Rob (\S 11), Des (\S 16)\\
& Others & $8$ & $0$\\
{\bf Subtotals} &  & $23$ & $4$\\ \midrule
Security & Any & $22$ & $0$\\
{\bf Subtotals} &  & $22$ & $0$\\ \midrule
Maintainability & Modularity & $4$ & $2$ & Tra (\S 17)\\
& Analysability & $6$ & $1$ & Tra (\S 20)\\
& Modifiability & $7$ & $1$ & Des (\S 18)\\
& Testability & $6$ & $1$ & Rob (\S 14)\\
& Others & $4$ & $0$\\
{\bf Subtotals} &  & $27$ & $5$\\ \midrule
Portability & Adaptability & $6$ & $1$ & Req (\S 1)\\
& Replaceability & $8$ & $2$ & Req (\S 7)\\
& Others & $5$ & $0$\\
{\bf Subtotals} &  & $19$ & $3$\\ \midrule \midrule
{\bf Total} & & $179$ & $15$\\ \bottomrule 
\end{tabular}
\end{table*}
\begin{table}[h]
\caption{Impact analysis on Data quality model} \label{data_q}
\begin{tabular}{@{} lrrl @{}} \toprule
& \multicolumn{3}{c}{Number of QMEs} \\ \cmidrule(r){2-4}
Characteristic & all & \multicolumn{2}{c}{affected} \\ \midrule
Accuracy & $14$ & $2$ & Req $\times 2$ (\S 2, \S 8), Rob (\S 15)\\
Completeness & $16$ & $0$\\
Consistency & $12$ & $0$\\
Credibility & $8$ & $0$\\
Currentness & $6$ & $0$\\
Accessibility & $6$ & $0$\\
Compliance  & $4$ & $0$\\
Confidentiality & $4$ & $0$\\
Efficiency & $14$ & $0$\\
Precision & $4$ & $0$\\
Traceability & $6$ & $0$\\
Understandability & $14$ & $0$\\
Availability & $6$ & $0$\\
Portability & $6$ & $0$\\
Recoverability & $6$ & $0$\\ \midrule \midrule
{\bf Total} & $126$ & $3$\\ \bottomrule 
\end{tabular}
\end{table}
\begin{table}[h]
   \small
\caption{Impact analysis on Quality in use model} \label{use_q}
\begin{tabular}{@{} llrrl @{}} \toprule
& & \multicolumn{3}{c}{Number of QMEs} \\ \cmidrule(r){3-5}
Characteristic & Sub-characteristic & all & \multicolumn{2}{c}{affected} \\ \midrule
Effectiveness & Any & $8$ & $0$\\
{\bf Subtotals} &  & $8$ & $0$\\ \midrule 
Efficiency & Any & $11$ & $0$\\
{\bf Subtotals} &  & $11$ & $0$\\ \midrule 
Satisfaction & Any & $13$ & $0$\\
{\bf Subtotals} &  & $13$ & $0$\\ \midrule 
Freedom from risk & Any & $21$ & $0$\\
{\bf Subtotals} &  & $21$ & $0$\\ \midrule 
Context coverage & Context completeness & $2$ & $1$ & Req (\S 3)\\
 & Flexibility & $6$ & $1$ & Req (\S 4)\\
{\bf Subtotals} &  & $8$ & $2$\\ \midrule \midrule
{\bf Total} & & $61$ & $2$\\ \bottomrule 
\end{tabular}
\end{table}

The characteristics of machine learning models that affected QMEs the most were 
a lack of requirements specification and a lack of robustness.
First, we discuss the impact of a lack of requirements specification. 
Quality characteristics involving preconditions (such as operational contexts, 
the interval of values, and operational environments) were affected by a lack of requirements specification. 
This is because requirements specifications define preconditions for systems. 
As discussed previously, machine learning models are trained using data-driven processes and lack explicit requirements specifications. 
Instead, preconditions are implicitly encoded in training data and not explicitly described. 
Thus, the following QMEs become unmeasurable due to a lack of preconditions (requirements specifications).
\begin{description}
\item[\S 1] Number of functions which were tested in different operational environments \\  \relax {\small [Product quality / Portability / Adaptability / Operational environment adaptability]}
\item[\S 2] Number of data items for which can be defined a required interval of values \\ \relax {\small [Data quality / Accuracy / Data accuracy range]}
\item[\S 3] Total number of required distinct contexts of use \\ \relax {\small [Quality in use / Context coverage / Context completeness / Context completeness]}
\item[\S 4] Total number of additional contexts in which the product might be used \\ \relax {\small [Quality in use / Context coverage / Flexibility / Flexible context of use]}
\end{description}
Note that the corresponding quality model, characteristic, sub-characteristic, and QM are described in square brackets.
Different operational environments in which systems must be tested, required intervals of values for data items, distinct contexts of use, 
and additional contexts in which the product might be used cannot be defined for data-driven training processes; 
the above QMEs are not measurable.

The reasons for a lack of requirements specifications in machine learning models are twofold: 
a lack of preconditions (introduced in the last paragraph) 
and a difficulty defining the desired behaviors of machine learning models due to the wide variety of input and output patterns.
For example, there are numerous variations of pedestrians (such as young and old, one with bags and umbrella) for an AEB function 
and it is difficult to define 
the function precisely (the types of pedestrians that the system covers) without ambiguity.
Being unable to define precise functions affects Function suitability, as well as Portability of Product quality. 
Being unable to define precise normal conditions, outliers for a wide variety of input data values are not definable, neither.
The following QMEs are not measurable due to the difficulty of defining behaviors:
\begin{description}
\item[\S 5] Number of functions that are incorrect \\ \relax {\small [Product quality / Functional suitability / Functional correctness / Functional correctness]}
\item[\S 6] Number of functions missing or incorrect among those that are required for achieving a specific usage objective \\ \relax {\small [Product quality / Functional suitability / Functional appropriateness / Functional appropriateness of usage objective]}
\item[\S 7] Number of functions which produce similar results as before \\ \relax {\small [Product quality / Portability / Replaceability / Functional inclusiveness]}
\item[\S 8] Number of data values that are outliers \\ \relax {\small [Data quality model / Accuracy / Risk of data set inaccuracy]}
\end{description}

Next, we discuss the impact of a lack of robustness. 
QMEs that observe machine learning system behavior are affected by a lack of robustness. 
When the inputs of machine learning models change even slightly, the results can change drastically. 
Therefore, the behavior of such systems becomes uncertain and we cannot measure (count) correct behavior. 
Moreover, we noticed that the QMEs affected by low robustness were similar to those affected by a lack of requirements specification. 
The QMs using these QMEs are typically ratios, with numerators being QMEs that count correct behavior 
and denominators being QMEs that count preconditions.
For example, one of the quality measures of Functional correctness is $X = 1 - A/B$, 
where $A = \text{[Number of functions that are incorrect]}$, $B = \text{[Number of functions considered]}$. 
We cannot measure the numerator $A$ and the denominator $B$ due to the two characteristics of 
machine learning models --- a lack of robustness and a lack of requirements specification, respectively.
The following QMEs are not precisely measurable due to a lack of robustness:
\begin{description}
\item[\S 9] Number of functions that are incorrect \\ \relax {\small [Product quality / Functional suitability / Functional correctness / Functional correctness]}
\item[\S 10] Number of functions missing or incorrect among those that are required for achieving a specific usage objective \\ \relax {\small [Product quality / Functional suitability / Functional appropriateness / Functional appropriateness of usage objective]}
\item[\S 11] Number of avoided critical and serious failure occurrences based on test cases \\ \relax {\small [Product quality / Reliability / Fault tolerance / Failure avoidance]}
\end{description}

QMEs related to negative events affected the difficulty of capturing rare cases of machine learning models, 
which is another form of a lack of robustness. 
Outliers and failures in SQuARE should have included rare cases; however, rare cases may not appear in a limited time frame 
and when they do, the extremely low probability of occurrence may be neglected. 
As mentioned previously, an extremely long FOT is required to capture such rare events. 
The following are the QMEs that were underestimated due to the difficulty of overlooking rare cases:
\begin{description}
\item[\S 12] Number of system/software failures actually occurred \\ \relax {\small [Product quality / Reliability / Maturity / Mean time between failure, MTBF]}
\item[\S 13] Number of failures detected during observation time \\ \relax {\small [Product quality / Reliability / Maturity / Failure rate]}
\item[\S 14] Number of test functions required \\ \relax {\small [Product quality / Maintainability / Testability / Test function completeness]}
\item[\S 15] Number of data values that are outliers \\ \relax {\small [Data quality / Accuracy / Risk of data set inaccuracy]}
\end{description}

There is a small impact on machine learning systems due to the lack of design specification and lack of transparency characteristics. 
If there are no design specifications, we cannot estimate the effort of a system modification nor the impact of a local modification 
to the overall system. 
We cannot forecast how many hours the training process will require, in advance. 
In addition, we cannot know the strengths and weaknesses of automatically trained machine learning models in general. 
Therefore, we cannot know the redundancy of components without design specification or transparency. 
Models with similar weaknesses do not work as redundancies, 
and redundant installation does not make sense for machine learning models.
The following are QMEs that are unmeasurable due to a lack of design specifications and a lack of transparency:
\begin{description}
\item[\S 16] Number of system components redundantly installed \\ \relax {\small [Product quality / Reliability / Fault tolerance / Redundancy of components]}
\item[\S 17] Number of components which are implemented with no impact on others \\ \relax {\small [Product quality / Maintainability / Modularity / Coupling of components]}
\item[\S 18] Expected time for making a specific type of modification \\ \relax {\small [Product quality / Maintainability / Modifiability / Modification efficiency]}
\end{description}
Since there is no established method of diagnostic and monitoring functionalities for machine learning models, the following QMEs are not measurable for machine learning systems.
\begin{description}
\item[\S 19] Number of functions having state monitoring capability \\ \relax {\small [Product quality / Usability / Operability / Monitoring capability]}
\item[\S 20] Number of diagnostic functions useful for causal analysis \\ \relax {\small [Product quality / Maintainability / Analysability / Diagnosis function effectiveness]}
\end{description}

We have discussed the combination of machine learning characteristics 
with a conventional system and software quality standard, SQuARE.
The typical gaps for the quality models of machine learning systems were found in requirements specification 
(precondition specification and level of detail for function specification) and 
robustness (uncertainty of observation and extremely low probable rare cases). 
In order to address these gaps, system quality models can be modified and/or extended.
We introduce the direction to address these gaps in \ref{ml_model}.

\subsection{Toward quality models for machine learning systems}
\label{ml_model}
The first set of challenges exist in quality measures for preconditions and functions (functionalities) for 
machine learning systems, that is, requirements specification.
We assume that preconditions and function specifications are defined by input range and pairs of input/output, respectively. 
If input and/or output data are high-dimensional, both defining preconditions and detailed function specifications are difficult.
As machine learning models are trained in a data-driven manner, we inevitably conclude that data is involved.
One natural idea is first to manually engineer the deductive specifications in as detailed a manner as possible and second to prepare data 
that includes example instances for requirements specifications.
Requirements specifications of machine learning systems cannot fully define the preconditions and functions; 
however, the remaining uncertainty of specifications is covered by examples.
In order to make requirements specifications as detailed as possible, 
we need quality definitions (and subsequently QM) of requirements specifications themselves.
A type of QM for requirements specifications is the sum of the quality of deductive requirements specification and 
the quality of inductive requirements specification, that is, sample data.

The quality of deductive requirements specifications for 
machine learning systems --- that is, the level of details of requirements specifications --- is not straightforward to measure.
Although not quantitative, a proxy of quality of deductive requirements specification is 
to measure the level of detail of the background argument. 
An earlier study~\cite{DBLP:conf/safecomp/IshikawaM18} used structured arguments, 
like goal structure notation (GSN)~\cite{Kelly04thegoal}, 
to address uncertain requirements and environments.
Quality measures of deductive requirements specification such as the following 
can be added to quality models of machine learning systems:
\begin{itemize}
\item "Number of functions with preconditions specified with structured argument" divided by "Number of functions that could benefit from specifying preconditions"
\item "Number of functions with detailed function specification with structured argument" divided by "Number of functions that could benefit from detailed function specification"
\end{itemize}
Further, the quality of inductive requirements specification (sample data) must be defined as 
the coverage of deductive requirements specifications, 
that is, how much deductive requirements specifications were covered by the inductive requirements specification (sample data).
If the structured argument is in a tree structure, the ratio of leaf nodes that have corresponding sample data can be 
a quality measure of sample data.
A quality measure of inductive requirements specification is given by the following:
\begin{itemize}
\item "Number of GSN solutions (leaf nodes) having corresponding sample data" divided by "Number of GSN solutions (leaf nodes) that could benefit from specifying sample data"
\end{itemize}

It is also important to handle the uncertainty of observation of machine learning systems in the quality models for machine learning systems.
The current quality measures are deterministic. Introducing a number of trials and variance to quality measures 
will incorporate the uncertainty of observation and improve the expression power of quality models.

Another aspect in the lack of robustness is the extremely low probability of rare cases. It must be noted that it is not possible to identify all rare cases by definition. 
We cannot evaluate the result of rare case discovery; however, we can see the quality of the process or the effort involved in it.
Quality measures of the process of discovering rare cases would be 
the effort in rare case discovery plus the number of rare cases discovered in a unit of time.

A viewpoint that is not included in the current quality models is development data, although SQuARE has Data quality.
Data quality in SQuARE is about the data included in the system itself, such as customer mailing address database.
For machine learning systems, development data --- that is, test and training data --- is rather important, 
and the quality models for machine learning systems must include the corresponding quality.
There are two different definitions for test data quality and training data quality.
The former is related to inductive requirements specification, that is, an aforementioned quality of sample data. 
Test data quality includes the gap between manually engineered deductive requirements specifications 
and actually collected sample data points.
The latter is related to design specification and includes the quality of manually annotated supervisory signals. 
This will be a trade-off to the cost of labor-intensive annotation processes.

\section{Conclusion}
With the rapid development of technology in recent years, machine learning has begun to be employed in various systems. 
To use machine learning in a safety-critical system, such as an automated driving system, 
it is necessary to demonstrate the safety and security of the system to society through the engineering process. 
In this paper, taking automated driving as an example, 
we presented open engineering problems with corresponding related works and research directions 
from the viewpoints of requirements, designs, and verifications for machine learning models and systems. 

At the level of the machine learning model, 
we hypothesized an ideal training process that connects deductive requirements and data-driven training, 
thereby considering test data as a requirements specification and training data as a design specification. 
Moreover, we recognized that the characteristics of machine learning models are a lack of requirements specification, 
a lack of design specification, a lack of interpretability, and a lack of robustness.
We also discussed the combination of a conventional system and software quality standard, SQuARE, 
and the aforementioned characteristics of machine learning models to study the quality models for machine learning systems. 
It turned out that a lack of requirements specification (precondition specification and level of detail for function specification) and 
a lack of robustness (uncertainty of observation and extremely low probability rare cases) have the largest impact 
on the conventional system quality models.
Further, we discussed the direction of future quality models for machine learning systems; 
however, most of it is a subject for future research.

Future research directions include the development of element technologies for engineering machine learning models and systems, 
such as requirements specification techniques to cover test data distribution or open environments. 
As evident from this paper, there are numerous open engineering problems and possible directions to address them.
However, to establish an engineering process for safety-critical machine learning systems, 
even if each company individually performs its own engineering processes based on its own concepts, 
process activities and work products cannot be automatically accepted by human society. 
Individual practices are not standard, and in order to achieve accountability, 
need evaluation on a case-by-case basis by a third party, particularly in case of problems. 
In such evaluations, own engineering practices of individual companies are at risk for being misunderstood, 
otherwise proprietary development information has to be disclosed for accountability. 
Thus, we need widely accepted standards to avoid these situations.
Attempts to research element technologies along with standard guidelines for requirements, 
designs, and verifications would also be practically helpful.
For example, a standard guideline for multiple verification tiers 
(actual data testing for normal conditions, simulated data testing for the corner cases, 
automatic verification for highest integrity levels only, falsification in middle integrity levels, etc.) 
would encourage the practical use of verification techniques 
and help an industry suffering from a lack of quality assurance of machine learning systems.
Another approach is to develop standard quality models for machine learning systems. 
In this paper, we discussed the quality models for machine learning systems based on SQuARE. 
Future research directions include discussing quality characteristics beyond SQuARE, 
defining specific QM and QME, and quality characteristics and sub-characteristics if necessary.

\noindent
\\
Conflict of Interest: The authors declare that they have no conflict of interest.

\bibliographystyle{spbasic}
\bibliography{ref}

\end{document}